\documentclass[11pt,a4paper]{article}

\pdfoutput=1
\usepackage{jheppub}
\usepackage{amsmath,bbm}
\usepackage{amssymb}
\usepackage[normalem]{ulem}
\usepackage{graphicx} 
\graphicspath{{figures/}}
\usepackage[latin1]{inputenc}
\usepackage{mathrsfs}
\usepackage{mathtools}
\usepackage[dvipsnames]{xcolor}
\usepackage{xcolor}
\usepackage{hyperref}
\usepackage{upgreek}

    \makeatletter
    \let\@fnsymbol\@arabic
    \makeatother

\newcommand{\bea}{\begin{eqnarray}}
\newcommand{\eea}{\end{eqnarray}}
\newcommand{\be}{\begin{equation}}
\newcommand{\ee}{\end{equation}}
\newcommand{\ba}{\begin{array}}
\newcommand{\ea}{\end{array}}

\def\gsim{\raise0.3ex\hbox{$\;>$\kern-0.75em\raise-1.1ex\hbox{$\sim\;$}}}
\def\lsim{\raise0.3ex\hbox{$\;<$\kern-0.75em\raise-1.1ex\hbox{$\sim\;$}}}

\title{
	Exotic Higgs decays into displaced jets at the LHeC
	}
	
\preprint{\begin{flushright} APCTP-Pre2020-018  \\  LTH 1242 \\ TTP20-030
\end{flushright}}	
	
\author[a,b,c]{Kingman Cheung}
\author[d,e]{Oliver Fischer}
\author[c,f]{Zeren Simon Wang}
\author[g,h]{Jose Zurita}
\affiliation[a]{\it Department of Physics, Konkuk University, Seoul 05029, Republic of Korea}
\affiliation[b]{\it Physics Division, National Center for Theoretical Sciences, Hsinchu, Taiwan} 
\affiliation[c]{\it Department of Physics, National Tsing Hua University, Hsinchu 300, Taiwan}
\affiliation[d]{\it Max-Planck-Institut f\"{u}r Kernphysik, 69117 Heidelberg, Germany} 
\affiliation[e]{\it Department of Mathematical Sciences, University of Liverpool, Liverpool, L69 7ZL, UK}
\affiliation[f]{\it Asia Pacific Center for Theoretical Physics (APCTP) - Headquarters San 31, Hyoja-dong, Nam-gu, Pohang 790-784, Korea}
\affiliation[g]{\it Institute for Nuclear Physics (IKP), Karlsruhe Institute of Technology, Hermann-von-Helmholtz-Platz 1, D-76344 Eggenstein-Leopoldshafen, Germany}
\affiliation[h]{\it Institute for Theoretical Particle Physics (TTP), Karlsruhe Institute of Technology, Engesserstra{\ss}e 7, D-76128 Karlsruhe, Germany}
\emailAdd{cheung@phys.nthu.edu.tw}
\emailAdd{oliver.fischer@liverpool.ac.uk}
\emailAdd{wzs@mx.nthu.edu.tw}
\emailAdd{jose.zurita@kit.edu}

\date{\today}

\abstract{
Profiling the Higgs boson requires the study of its non-standard decay modes.
In this work we discuss the prospects of the Large Hadron electron Collider (LHeC) to detect scalar particles with masses $\gtrsim$ 10 GeV produced from decays of the Standard Model (SM) Higgs boson.
These scalar particles decay mainly to bottom pairs, and in a vast portion of the allowed parameter space they acquire a macroscopic lifetime, hence giving rise to displaced hadronic vertices.
The LHeC provides a very clean environment that allows for easy identification of these final states, in contrast to hadronic colliders where the overwhelming backgrounds and high pile-up render such searches incredibly challenging.
We find that the LHeC provides a unique window of opportunity to detect scalar particles with masses between 10 and 30 GeV.
In the Higgs Portal scenarios we can test the mixing angle squared, $\sin^2 \alpha$, as low as $10^{-5} - 10^{-7}$, with the exact value depending on the vacuum expectation value of the new scalar. 

Our results are also presented in a model-independent fashion in the lifetime-branching ratio and mass-branching ratio planes.
We have found that exotic branching ratios of the Higgs boson at the sub-percent level can be probed, for the scalar decay length in the range
$10^{-4}$ m $\lesssim c \tau \lesssim 10^{-1}$ m.
The expected coverage of the parameter space largely exceeds the published sensitivity of the indirect reach at the high-luminosity Large Hadron Collider via the invisible Higgs branching ratio. 

}

\notoc 
\begin{document}

\maketitle

\section{Introduction}
\label{sec:intro}

We are now entering the era of precision measurements of the 125 GeV Higgs boson \cite{Aad:2015zhl,Khachatryan:2016vau}, which was first
discovered in 2012 \cite{Aad:2012tfa,Chatrchyan:2012ufa}.
All data are best described by the Standard Model (SM) Higgs boson (see e.g. ref.~\cite{Cheung:2018ave}). 
Even though there is no sign of physics beyond the Standard Model (BSM) at the Large Hadron Collider (LHC), the open questions of particle physics remain, such as the
underlying mechanism of electroweak symmetry breaking, the nature of dark matter, the origin of neutrino masses, and many others.

The null result from the BSM searches at the LHC raises the question if there is a systemic shortcoming.
Indeed, new physics may manifest itself in the form of long-lived particles (LLP), which might have escaped from detection up to now because previous efforts (including analyses, and hardware and software triggers at CMS and ATLAS) focused on {\it promptly decaying new particles,} such as squarks or gluinos in supersymmetry frameworks, and top partners in composite models.
Nowadays, there is a rising interest in searches for LLPs in both the theoretical and experimental communities~\cite{Alimena:2019zri}.

From the theoretical point of view, many BSM models naturally predict the existence of LLPs. 
A classical example is the breaking of supersymmetry in a hidden sector, which typically interacts with the SM via gravity or new gauge groups.
Models with hidden sectors employ a new (often unbroken) symmetry that requires the interactions with the SM to be mediated by so-called {\it portal} fields.
Notable examples of such hidden sector theories include Higgs-portal models~\cite{Silveira:1985rk}, gauge-mediated supersymmetric models~\cite{Dimopoulos:1996vz}, hidden-valley models~\cite{Strassler:2006im}, and neutral-naturalness scenarios~\cite{Chacko:2005pe,Burdman:2006tz}.

Here, we focus on the so-called ``Higgs portal'' or ``scalar portal'', where the SM Higgs field is coupled to additional scalar degrees of freedom 
in the hidden sector via renormalizable couplings. 
In these scalar portal models the CP-even neutral degrees of freedom mix with the SM Higgs boson to form additional mass eigenstates.
The heavy ones, referred to as ``heavy'' or ``exotic'' Higgses, are being searched for at the LHC, see for instance ref.~\cite{Adhikary:2018ise} and references therein, while mass eigenstates with masses below half of the SM Higgs mass can be pair produced from and searched for in Higgs boson decays.
In general, such scalar bosons can be observed via their prompt decays into SM particles when heavy \cite{Aaboud:2018iil}, or via the exotic effects from their possibly long lifetime \cite{Curtin:2013fra}. 
For masses around a few GeV these scalar bosons can be tested at low energy experiments, cf.\ ref.~\cite{Boiarska:2019jym}.

The classes of signatures that are connected to LLPs depend on their lifetime and decay products. 
Easiest to access and study are the all-leptonic decay modes, where LLPs decay exclusively into charged leptons to form displaced leptons or lepton jets~\cite{ArkaniHamed:2008qp}, depending on the LLP boost. 
More arduous from the analysis point of view are all-hadronic decay modes, including emerging jets~\cite{Schwaller:2015gea}, dark jets \cite{Park:2017rfb}, and semi-visible jets~\cite{Cohen:2015toa}, depending on the fraction of invisible, i.e.\ uncharged, particles in the LLP decays.

Studying LLP signatures at the LHC is a challenging task at best, seeing that the existing triggers are optimized for prompt decays of heavy particles, and these specific BSM signals might have been missed. 
Recently, the LHC collaborations have developed a broad program of LLP searches, cf.\ e.g.\ ref.~\cite{Alimena:2019zri}.
Specific triggers will be installed in future runs at the ATLAS and CMS experiments \cite{Aad:2013txa,CMS:2018qgk,Bhattacherjee:2020nno} to ameliorate this shortcoming, and external detectors are proposed to improve detection prospects \cite{Barron:2020kfo}.
In general, however, the QCD rich environment in proton-proton collisions at high pile-up rates is not an ideal place to search for certain classes of LLPs.

Better prospects of finding not-so-heavy particles with macroscopic lifetimes can be expected at the presently discussed Large Hadron electron Collider (LHeC) \cite{Klein:2009qt,AbelleiraFernandez:2012cc, Bruening:2013bga,Agostini:2020fmq}, which foresees the construction of a new 60 GeV electron beam, to be collided with one of the LHC's 7 TeV proton beams, which would result in $\sim 1.3$ TeV center-of-mass energy with large luminosities, planned to run concurrently with the High-Luminosity  LHC (HL-LHC).
The clean (read: no pile-up and low QCD rates) environment and the excellent tracking resolution are quality features for all BSM studies where particles of small mass and subsequently soft decay products are difficult to study at the LHC \cite{Curtin:2017bxr}. For an overview of BSM studies at hadron-electron colliders, we refer the reader to Chapter 8 of reference~\cite{Agostini:2020fmq}.

In this article, we discuss the prospects to search for additional scalar bosons with masses below half the Higgs mass at the LHeC, produced from the decays of the SM Higgs boson, which at the LHeC is produced via vector-boson fusion (VBF).
While we focus our analysis on light scalars with masses above 10 GeV and decaying into a pair of bottom quarks, we like to note that lighter masses are also of interest, see for example refs.~\cite{Chang:2016lfq, Winkler:2018qyg,CidVidal:2019urm,Cheung:2019qdr,Alipour-Fard:2018lsf}. 
In this article we go beyond simple geometrical cuts employed in previous works, by carrying out the analysis with detailed event simulations including detector effects in order to make refined estimates of the experimental acceptance and efficiencies.

The article is structured as follows. In section~\ref{sec:model}, we highlight the details of our theoretical setup, including our model of choice and specifying our model-independent framework. In section~\ref{sec:setup}, we give details of the analysis and search strategies, emphasizing the need for realistic simulations.
In section~\ref{sec:resu}, we show our numerical results, and we reserve section~\ref{sec:conclu} for our conclusions.

\section{Theoretical setup and experimental constraints}
\label{sec:model}

The goal of this work is to obtain the sensitivity of the future LHeC to scalar LLPs from their decays to displaced jets.
For convenience of illustration we adopt a simplified working model to carry out our analysis, with three model parameters that control the LLP's production rate and their decay length.
In order to make our study more useful for more generic scalar LLP models, we also express our result directly as a function of the production rate (here the exotic Higgs boson decay branching ratio, as the SM Higgs production rate is fixed), mass of $h_2$, and its lifetime (or decay length), which allows for straightforward reinterpretations \cite{Abdallah:2020pec}.

We consider the simple model that was already employed to survey the potential of the LHeC searching for new (heavy) scalars \cite{DelleRose:2018ndz}. 
The SM is extended with one complex neutral scalar field $S$ that is a singlet under the SM gauge group.  
In this model, the scalar sector consists of a new singlet field $S$ and the Higgs doublet $H$, and it is described by the potential
\bea
\label{eq:scalarpotential}
V(H,S) = - \mu_1^2 H^\dag H - \mu_2^2 \, S^\dag S + \lambda_1 (H^\dag H)^2 + \lambda_2 (S^\dag S)^2 + \lambda_3 (H^\dag H)(S^\dag S)\,.
\eea
The above is the most general renormalizable scalar potential of the SM $SU(2)$ Higgs doublet $H$ and the complex scalar $S$.
After electroweak symmetry breaking, both the SM Higgs doublet field $H$
and the new scalar singlet field $S$ develop vacuum expectation values (vevs)
$\langle H \rangle \equiv v/\sqrt{2} $ and
$\langle S \rangle \equiv x /\sqrt{2}$, respectively,
with $v \approx 246$ GeV. 
Using the tadpole conditions $\partial V / \partial H=0$ and
$\partial V / \partial S=0$, we can express $\langle H \rangle$
and $\langle S \rangle $ as
\begin{eqnarray}
\langle H \rangle^2 \equiv \frac{v^2}{2} & = &  \frac{1}{2} \left(
  \frac{ 4 \lambda_2 \mu_1^2 - 2 \lambda_3 \mu_2^2 }
       { 4 \lambda_1 \lambda_2 -  \lambda_3^2  } \right ), \nonumber \\
\langle S \rangle^2 \equiv \frac{x^2}{2} & = &
\frac{1}{2} \left(
  \frac{ 4 \lambda_1 \mu_2^2 - 2 \lambda_3 \mu_1^2 }
       { 4 \lambda_1 \lambda_2 -  \lambda_3^2  } \right ).
\end{eqnarray}
We then expand around the vacua of $H$ and $S$ to define the physical fields with
\begin{equation}
H = \frac{v + H_r +i H_i} {\sqrt{2}}\,, \qquad
S = \frac{x + S_r +i S_i} {\sqrt{2}}\,,
\end{equation}
where ``$r$'' and ``$i$'' in the subscript denote the real and imaginary
components of the fields.
The mixing term proportional to $\lambda_3$ induces the mixing between the $H$ and $S$. 
The mass eigenstates from the resulting mass matrix correspond to the physical fields through a mixing angle
\bea
\left( \begin{array}{c} h_1 \\ h_2 \end{array} \right)
= \left( \begin{array}{cc} \cos \alpha & - \sin \alpha \\
 \sin \alpha & \cos \alpha \end{array} \right)  \left( \begin{array}{c} H_r
\\ S_r \end{array} \right).
\eea
After mixing we identify $h_1$ with the observed 125 GeV Higgs boson.

We can swap the parameters in the Lagrangian~\eqref{eq:scalarpotential} for the physical masses, the scalar mixing angle $\alpha$, and the vacuum expectation values $v$ and $x$ of the $H$ and $S$ fields, respectively. 
Since in our setup $m_{h_1}$ and $v$ are known, our three unknowns are $m_{h_2}$, $\alpha$, and $x$.
To leading order in the mixing angle $\alpha$, we can express the parameters as
\begin{eqnarray}
m_{h_1}^2 & \simeq & 2 \lambda_1 v^2 \, , \\
m_{h_2}^2 & \simeq & 2 \lambda_2 x^2 \, ,\\
\alpha & \simeq &- \frac{ \lambda_3 v x}{ m_{h_1}^2  - m_{h_2}^2 } \, , \\
{\cal L}_{h_1 h_2 h_2} & \simeq & - \frac{ \lambda_3 v }{2} h_1 h_2 h_2 \, .
\end{eqnarray}
We note that the small mixing limit is already favored by existing LHC data~\cite{ATLAS:2020cjb}, with $\sin^2 \alpha \lesssim 10^{-3}$ in the vast majority of the parameter space. We will discuss these limits in some detail at the end of this section.

We remark that the partial decay widths of $h_1$ into the SM channels 
are proportional to $\cos^2\alpha$, such that  the properties of the SM Higgs boson are recovered in the limit of ${\sin{\alpha} \to 0}$.
Also, there should be an additional physical mass eigenstate due to the imaginary part of the $S$ field, which we assume to be irrelevant\footnote{In $B-L$ theories, for instance, this degree of freedom is absorbed by the additional gauge boson in order to account for its mass. Alternatively, one can also choose a real scalar $S$ field, see e.g.\ ref.~\cite{OConnell:2006rsp}.}
in the rest of the paper.

The last term in eq.~\eqref{eq:scalarpotential} proportional to $\lambda_3$ gives rise to not only the mixing, but also to a coupling between $h_1$ and $h_2$, which yields,  e.g.~the additional decay channel $h_1 \rightarrow h_2 h_2$ if $m_{h_1}>2m_{h_2}$. 
We calculate the partial decay width of the SM-like Higgs boson $h_1$ into two scalars $h_2$:
\begin{equation} 
\Gamma (h_1 \to h_2 h_2 )
\simeq \frac{1} {32 \pi m_{h_1} } (\lambda_3 v)^2 \left (
 1 -  \frac{4 m_{h_2}^2}{m_{h_1}^2 } \right )^{1/2}
 \simeq \frac{\sin^2\alpha ( m_{h_1}^2 - m_{h_2}^2 )^2}{32 \pi m_{h_1}x^2 }
 \left (  1 - \frac{4 m_{h_2}^2}{m_{h_1}^2 } \right )^{1/2}.
\label{eq:Higgs_branching}
\end{equation}
The corresponding decay branching ratio is obtained simply by dividing this partial decay width by the total width of the Higgs boson $\Gamma_{\rm SM\, Higgs} \simeq 4.1$ MeV \cite{CERNHiggs}.

The scalar mixing yields tree-level couplings between the mass eigenstate $h_2$ and the SM fields that are proportional to those of a SM Higgs boson times $\sin{\alpha}$.  
This allows the $h_2$ to decay into a pair of SM fermions $f\bar{f}$ via the Yukawa couplings $Y_f$, as long as it
is kinematically allowed.  
Consequently, the partial decay widths of the light scalar are functions of $m_{h_2}$ and $\sin^2{\alpha}$:
\begin{equation}
\label{pwidth}
\Gamma (h_2 \to f \bar f) = \frac{ N_C (Y_f \sin \alpha)^2 }{8 \pi} m_{h_2}
 \, \left ( 1- \frac{4 m_f^2}{m_{h_2}^2} \right )^{3/2}
  \;,
\end{equation}
where $N_C=3\,(1)$ labels the color factor of the fermion $f=q\, (\ell)$, and $Y_f \equiv g m^{\rm run}_f /(2 m_W)$. 
Here we use the running mass for $m_f^{\rm run}$ evaluated at the scale $m_{h_2}$ in the Yukawa coupling in order to account for the leading-log correction.
Note that the $m_f$ used in eq.~(\ref{pwidth}) is the pole mass for the kinematic factor. 
If only channels of SM fermions are considered, the total width is given by
\begin{equation}
\Gamma_{\rm tot} = \sum_{ f } \, \Gamma (h_2 \to f \bar f) \;.
\end{equation}
The loop-induced decays into a pair of gluons or photons are subdominant and can be neglected for the purposes of our discussion.
Notice, that in more complex models the total $h_2$ width could be increased by additional decays into hidden-sector particles.
In the setup considered here, for $ m_{h_2} \lesssim 60$ GeV the total width is dominated by the $b\bar b$ mode.
The scaling of the decay length versus the mixing angle and the mass of
$h_2$ is roughly given by 
\begin{equation}
\label{eq:ctau_HiggsPortal}
c \tau = \frac{c}{\Gamma_{\rm tot}} \approx 1.2 \times 10^{-5} \,
\left ( \frac{10^{-7}}{ \sin^2 \alpha} \right )\,
\left( \frac{ 10\;{\rm GeV} }{ m_{h_2} } \right ) \;\;\; {\rm m} \;.
\end{equation}
For the numerical analysis we include all the decay channels of $h_2$ using the program HDECAY 3.4 \cite{Djouadi:1997yw,Djouadi:2018xqq}, which comprises state-of-the-art radiative corrections. The input values used in our analysis for 
$\Gamma_{\rm SM} (h_2)$ and Br$(h_2 \to b \bar{b})$ are shown in Appendix~\ref{app:hwidths}.

We would like to comment now on the model dependence of the results. As mentioned there are three free parameters in this model, namely $\{
m_{h_2}, \alpha, x\}$. 
It is easy to see that we can trade $\alpha$ and $x$ for the phenomenologically relevant Br$(h_1 \to h_2 h_2)$ and $h_2$ lifetime ($c\tau$) parameters. 
As stressed at the beginning of this section, there are added values to model-independent studies. 
Hence we present our results for both the Higgs portal model ($\{ \alpha, x, m_{h_2} \}$) and for a model-independent formulation, in terms of  $m_{h_2}$, the exotic Higgs branching ratio into the two scalars, and the $h_2$ lifetime.

Before we close this section, we briefly discuss the current and future experimental efforts on constraining a light scalar which mixes
with the SM Higgs boson.  
The presently best constraints stem from LHCb for masses below $m_b \sim 5$ GeV, which exclude ${\sin^2{\alpha} > 10^{-5}}$ close to $m_{h_2} \sim m_b$, and from low energy experiments and astrophysics for masses below 1 GeV, see e.g.~ref.~\cite{Beacham:2019nyx} and references therein.  
For masses above 10 GeV, the current \emph{direct} limits were obtained at LEP \cite{Acciarri:1996um, Barate:2003sz,Schael:2006cr} and are only
at the order of $10^{-2}$ for $\sin^2{\alpha}$.  
Planned future experiments at the lifetime frontier are expected to test this class of models for mixings as small as $\sin^2{\alpha} \sim 10^{-13}$ but are limited by the bottom mass, cf.~ref.~\cite{Beacham:2019nyx}. 
Both ATLAS and CMS have conducted searches for exotic branching fractions of the SM Higgs, as well as direct searches for displaced jets. 
On the so called ``invisible" Higgs branching fraction, the current bound from ATLAS~\cite{ATLAS:2020cjb} (CMS~\cite{Sirunyan:2018owy}) is 13 (19)\%\footnote{Recently ref.~\cite{Ngairangbam:2020ksz}, 
explored the use of deep-learning methods based on low-level calorimeter data, finding that the current dataset could shrink the bound further to 4.3\%. This very important result needs to be scrutinized by the experimental collaborations.}
while the HL-LHC is expected to reduce this number to 2.5\% \cite{Cepeda:2019klc}. 
We note, however, that these studies are only indirect probes of the light scalars, and a putative excess will require to characterize the new physics signal elsewhere. 
In contrast, LHCb searches for displaced jets~\cite{Aaij:2016xmb,Aaij:2017mic} and tests these scalars directly, with current bounds between 2\% and 50\% and future prospects in the 0.02 - 2\% range~\cite{LHCb:2018hiv}. The weakest limit of 2\% are obtained for $c \tau$ outside the $10^{-3} - 10^{-1}$ m range, and for small masses. The corresponding ATLAS~\cite{Aad:2019xav} and CMS~\cite{CMS:2020idp} searches set competitive bounds for relatively large masses. 
The latter, having more luminosity, set the strongest bounds on the exotic branching fraction: $0.5-10\%$ for scalar masses above 40 GeV, in the $10^{-3}-5 \times 10^{-1}$ m lifetime range. 
We note in passing that the CMS analysis explicitly states the difficulty to test smaller masses. 
The reason for this is that the decay vertex has a lower track multiplicity and that the tracks are highly collimated, which strongly affects the efficiency of the reconstruction algorithms.%
\footnote{\label{footnote3}The CMS collaboration does present results for displaced light jets from the decays of a scalar boson with 15 GeV mass, which are two orders of magnitude weaker than those for the 40 and 55 GeV benchmarks. 
This bound suffers indeed from the low track multiplicity.
Since light jets are considered, it is, however, not affected by uncertainties from tertiary vertex reconstruction, which would be relevant for $b$-hadron final states.
}

It is the difficulty of LHC searches for LLP scalars with masses between $2 \cdot m_b$ and $\sim 30$ GeV that leads us to the LHeC which, by virtue of its tracking resolution and low pile-up~\cite{Agostini:2020fmq}, provides a unique window of opportunity for such LLP searches, especially for lifetimes below $10^{-3}$ m.
It is important to realise that the LHeC's potential has to be compared to the full HL-LHC run, since the two colliders will run concurrently. Moreover, the LHeC has a broad physics program including, among others, precision QCD, invaluable improvements of the PDF measurements, and precision measurements of several crucial SM properties.

\section{Analysis}
\label{sec:setup}

\begin{figure}[t]
\centering
\includegraphics[width=0.5\textwidth]{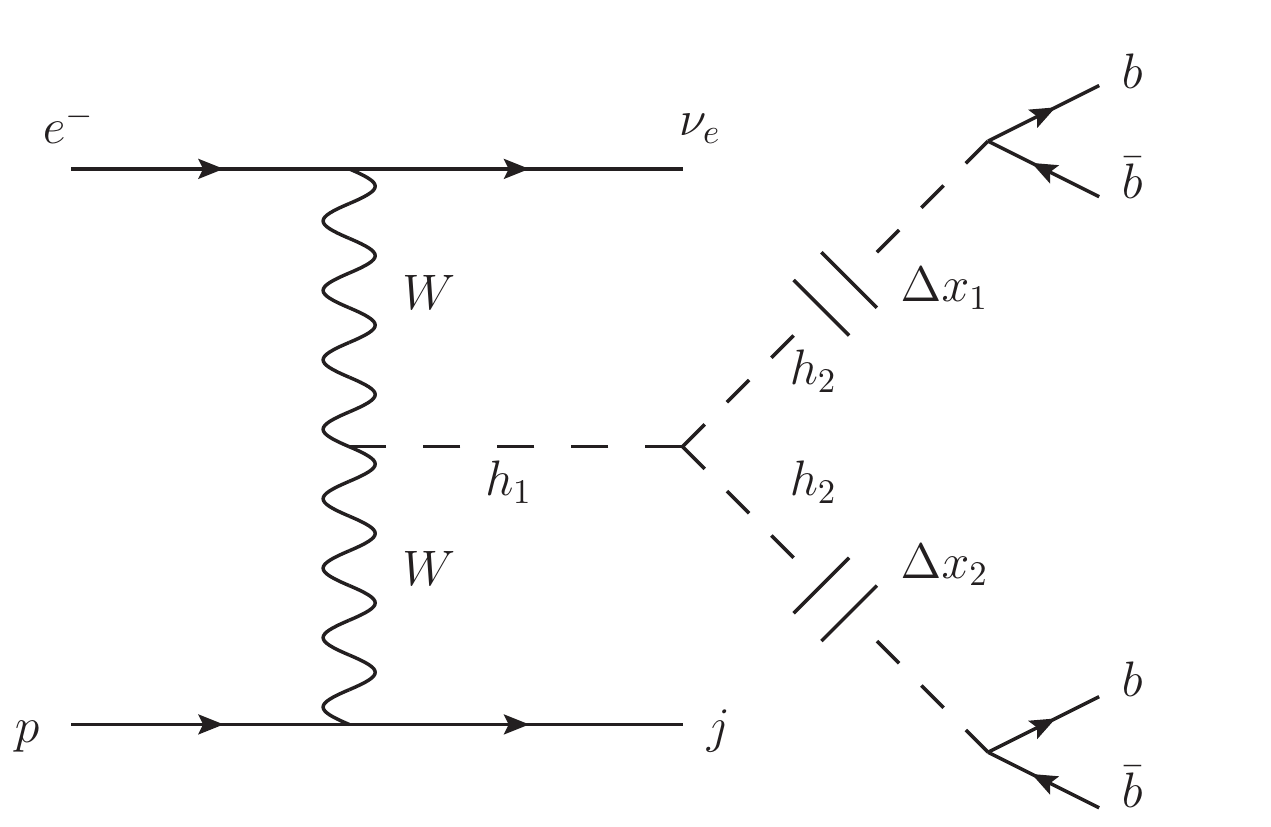}
	\caption{Charged-current VBF signature for long-lived $h_2$ search at proton-electron colliders. The light scalar bosons are pair produced from the decays of the SM Higgs boson $h_1$, and then decay with displacements $\Delta x_1$ and $\Delta x_2$, respectively, each into two $b-$jets.}
\label{fig:process}
\end{figure}

At proton-electron colliders such as the LHeC, the SM Higgs bosons are mainly produced in VBF processes via either $W$-bosons (CC: charged-current) or $Z$-bosons (NC: neutral-current). For the SM Higgs, NC production has a cross section approximately one order of magnitude smaller than the CC case~\cite{Agostini:2020fmq}.
Hence in this article we focus on the SM Higgs ($h_1$) CC production mode at the LHeC, with subsequent decays $h_1\rightarrow h_2 h_2$ and $h_2 \to b \bar{b}$, the latter decay being displaced\footnote{As discussed in the previous section, $h_2$ is naturally long-lived for a small enough mixing angle, which is strongly favored by current experimental constraints.}:
\begin{equation}
	p \, e^- \rightarrow \nu_e \, j \, h_1 \rightarrow \nu_e \, j \, h_2 \, h_2 \rightarrow \nu_e \, j \, (b\bar{b})_{\text{displaced}}\, (b\bar{b})_{\text{displaced}}. \label{eqn:process}
	\end{equation}
The signal process is depicted in figure \ref{fig:process}. 
As mentioned before, we consider $m_{h_2}$ between $10$ GeV and $m_{h_1}/2$, thus enforcing the on-shell decays of $h_1$ and $h_2$. We remark that we consider $h_2$ masses above the $b \bar{b}$ threshold for two reasons: first, the partonic picture breaks down for smaller masses and one must consider scalar decays into hadrons instead (cf.\ ref.~\cite{Winkler:2018qyg}) which affects the accuracy of our simulation setup; second, low energy experiments such as Belle and Belle II~\cite{Filimonova:2019tuy}, or DarkQuest~\cite{Batell:2020vqn} can be expected to have better sensitivity.

\subsection{Event Generation}
We use the parton-level Monte Carlo (MC) simulation tool MadGraph5\_aMC 3.0.2 with the Hidden Abelian Higgs model \cite{Curtin:2013fra,Curtin:2014cca} (HAHM)~\footnote{We note that the scalar sector of the HAHM is exactly equal to our equation~\ref{eq:scalarpotential}.}, to generate signal samples (each containing $10^5$ events) with the proton (electron) beam energy set at 7000 (60) GeV and without any beam polarization.
For the signal we perform a grid scan for $m_{h_2}$ between 10 GeV and $m_{h_1}/2$, and $c\tau$ between $10^{-12}$ m and 100 m. Concretely, we sampled every 2 GeV for 16 GeV $\leq m_{h_2} \leq$ 62 GeV, while we used a finer step size for the mass window 10 GeV $\leq m_{h_2} \leq$ 16 GeV. The scalar lifetime was scanned with a logarithmic step, using 57 points for the whole range considered here. 

For the simulation of the samples we implement the following parton-level cuts: $|p_T^{b/j}|>5$ GeV, $|\eta^{b/j}|<5.5$, and $\Delta$R$(b,b/j)>0.2$, where $p_T^{b/j}$ and $\eta^{b/j}$ denote the transverse momentum and pseudo-rapidity of the $b$/jet, and $\Delta$R$(b,b/j)$ is the angular separation of the $b$ quarks and jets.
The $p_T$ and angular separation thresholds are necessary in order to avoid the failing description of quantum chromodynamics (QCD) at low energies and/or for collinear emission, where the perturbation theory breaks down. The maximal value of the pseudo-rapidity is in accordance with the LHeC detector, which accepts particles with $-4.3 \leq \eta \leq 4.9$.
We perform showering and hadronization of our parton-level events with Pythia 6.4.28 \cite{Sjostrand:2006za} patched for $ep$ collider studies \cite{py6_patched}.%
\footnote{
This patch switches off some internal cuts relating to QCD processes in the very forward region of the beam remnant in order to enhance simulation efficiency. We remark that this patch exists only for Pythia 6.4.28 and not for e.g. Pythia 8.}

The inclusive Higgs production at the LHeC via charged-current VBF processes and unpolarized beams is estimated to be 110 fb \cite{Agostini:2020fmq}.
With an expected integrated luminosity of 1 ab$^{-1}$, this corresponds to approximately $1.1\times 10^5$ on-shell Higgs bosons produced.
We remark that the LHeC design foresees polarization of the electron beam of up to $-80\%$, which enhances the cross section of all interaction processes with weak gauge bosons by a factor up to two.
Since similarly enhanced cross sections can be expected for the main backgrounds, the signal-over-background ratio remains the same but the significance is increased. Hence the results presented in this article are conservative.

It is tantamount to take into account the reduction in the production cross section of the signature process arising from parton-level generator cuts on $p_T^{b/j}$, $\eta^{b/j}$, and $\Delta$R$(b,b/j)$, especially for small $m_{h_2}$.
We find for the considered mass range, the production cross section with the parton-level cuts is between $1\%$ and $24\%$ of the inclusive result for $m_{h_2}$ between 10 and 12 GeV, while for masses above that the reduction is even milder.
This is mainly because for lighter $h_2$, the angular separation of the bottom quarks and jets tends to be smaller. 

\subsection{Detector simulation}
At the detector level we use the fast detector simulation tool Delphes 3.3.2 \cite{deFavereau:2013fsa} with an LHeC-specific detector card.
This detector card accepts charged tracks with radial displacements up to about 15 cm, corresponding to the radial position of the fourth tracking layer of the innermost central tracking unit.

The jet clustering is performed with the package FastJet 3.1.3 \cite{Cacciari:2011ma,Cacciari:2005hq} using the anti-$k_t$ algorithm \cite{Cacciari:2008gp}. 
Identifying displaced jets is a notoriously difficult problem in Delphes 3 and a working module has not been officially implemented. 
However, in ref.~\cite{Nemevsek:2018bbt} a customized version of Delphes 3.3.2 \cite{delphes3_miha} was introduced, including additional modules that allow the definition of a displaced jet. 
More specifically, the transverse displacement of a jet $d_T(j)=\sqrt{d_x^2(j)+d_y^2(j)}$ is defined to be the minimum $d_T$ of all the tracks associated to the jet which are required to have a transverse momentum larger than a certain threshold. 

For concreteness, in this work we cluster jets with $\Delta$R  = 0.4, and we make use of the displaced jet modules setting  $\Delta$R$(\text{track}, j)<0.4$ and $p_T(\text{track})>1$ GeV~\footnote{
This threshold can in principle be lowered, as the LHeC would be able to identify softer tracks~\cite{Agostini:2020fmq}. For instance, in ref.~\cite{Curtin:2017bxr} thresholds as low as 50 to 400 MeV have been discussed.
}. Finally, for vertex smearing we employ the same resolution as ATLAS  \cite{Aad:2011zb,Aad:2011cxa}.

\subsection{Background processes}
The relevant background processes at the parton level can be classified by their final states
\begin{eqnarray}
	p + e^- \rightarrow \nu_e +  j + n_b \, b + n_\tau \, \tau + n_j \, j,
\end{eqnarray}
where $n_b$, $n_j$, and $n_\tau$ are the numbers of bottom quarks, light jets, and tau leptons, respectively. 
We note that we explicitly distinguish the beam jet, which recoils against the electron beam, from the other $n_j$  light jets in the final state. 
Hence, at the parton level, we are dealing with $2 \to 2 + N$ processes, where $N=n_b + n_j + n_{\tau}$. 
Given that the cross sections fall logarithmically with $N$, we restrict ourselves to $N \leq 4$. 
All the background event generation is performed using the same pipeline as the signal, except that we use the default Standard Model implementation instead of the HAHM model.

In table~\ref{tab:bgd} we provide a summary of the twelve considered background processes including $n_b$, $n_\tau$, $n_j$, and their cross sections at the LHeC.
\begin{table}[t]
\begin{center}
        \begin{tabular}{c|c|c|c|c|c|c}
		$B_i$ & $B_1$    &   $B_2$  & $B_3$ & $B_4$ &  $B_5$  &  $B_6$      \\
        \hline
		$n_b$     & 0	&	0	&	0	&	0	&	2	&	2\\ 
		$n_\tau$  & 0   &0 &0 &0 &0 &0  	\\
		$n_j$    & 1   & 2 &3  & 4& 0 & 1 \\
		$\sigma$ [pb] &  $2.00$E$2$ & $1.20$E$2$  & $6.74$E$1$  &$3.59$E$1$   &$4.10$E$-1$  &$4.68$E$-1$\\ 
	\hline\hline
		 & $B_7$    &   $B_8$  & $B_9$ & $B_{10}$ &  $B_{11}$  &  $B_{12}$      \\
        \hline
                $n_b$     & 2   &       4       &       0       &       0       &       0       &       0\\
                $n_\tau$  & 0   &0 &2 &2 &2 &4          \\
                $n_j$    & 2   & 0 &0   & 1 & 2 & 0 \\
		$\sigma$ [pb]  & $4.06$E$-1$ & $5.88$E$-4$ &  $2.90$E$-2$  & $2.03$E$-2$  & $1.08$E$-2$  & $4.76$E$-6$
        \end{tabular}
	\caption{Monte Carlo simulated background events and the cross sections at the LHeC. $n_b$, $n_\tau$, and $n_j$ label the number of $b$, $\tau$, and $j$ in the final states of each process, respectively. $\sigma$ gives the cross section of each background process in picobarn where the scientific notation is used.}
        \label{tab:bgd}
\end{center}
\end{table}
For the background processes $B_5-B_{12}$  the cross sections are small, and hence our simulation can be carried out with low statistical uncertainty. 
In contrast, the first four background processes, corresponding to QCD multi-jets, have very large cross sections and demand very large event samples. 
Usually, these processes do not give rise to displaced objects and only a tiny fraction is expected to form an irreducible background, thus it is tempting to ignore them straight away.
However, it is unclear \textit{a priori} whether or not the product of a large cross section and a tiny selection efficiency will yield an appreciable number of events. 
Indeed, as we show in the next section, it is these processes that yield the primary contributions to the total number of background events.

It is important to stress here that a proper treatment of the backgrounds above must include, for fixed $n_b$ and $n_\tau$, a jet merging procedure of all the different $n_j$ light jet multiplicities. 
However, a full merged calculation as required here has an outrageous computational cost\footnote{In our tests a 8-core i7 CPU required 1 hour to obtain 100 matched Monte Carlo events (at the reconstructed level) for processes $B_1 - B_4$ using MLM matching with $k_T$ jets. A major bottleneck for the matching is that the typical hard scale of the process is very low, since jets coming from light scalar decays tend to have low $p_T$.}. 
Hence we have decided to allow ourselves to include multiple event counting, in order to significantly speed up the background event generation by several orders of magnitude.

\subsection{Search strategy and cutflow}
Here we describe our cut-based search strategy. 
Recall that the signal consists of the exotic Higgs decay into two $h_2$,
each of which in turn decays into two $b-$jets, such that the parton-level final state is given by the beam jet, four $b-$jets, and missing energy from the electron neutrino.
Therefore, our first selection criterion is the requirement of at least 5 jets \emph{at the reconstructed level}, $n_J \geq 5$. 
We here use $n_J$ to explicitly distinguish this quantity from the number of parton level jets $n_j$ (which, moreover, does not explicitly include the beam jet). 

Since the $h_2$ is long-lived, its decay products appear at a displaced production vertex.\footnote{
We remark that the LHeC detector allows to measure both, longitudinal and transverse displacements with high precision. 
However, only the transverse coordinate of the primary vertex, where the electron and the incident proton scatter, can be inferred from the accurate knowledge of the transverse extensions of the beams and the measured beam jet.
The longitudinal extension of the beam bunches result in a less precise knowledge of the longitudinal coordinate of the primary vertex. 
Hence, we shall only consider the transverse displacement in the following.}
The spatial resolution of the LHeC detector is 10 $\mu$m \cite{Agostini:2020fmq} and we conservatively label a jet as ``displaced" if its transverse displacement is $d_T(J)>50$ $\mu$m.
Consequently, our second selection criterion is the requirement of at least one displaced jet ($n_{\text{disp.} J} > 0$).

In our signal event samples, the two displaced $b-$jets stemming from one $h_2$ ought to share their displacement coordinates (which could be washed out due to jet-clustering and/or detector effects).
Our next step is to identify two $b-$jets with similar displacement in order to form $h_2$ candidates.
The displaced jets are grouped together into a so-called ``heavy group'' if the absolute difference of their respective transverse displacements is smaller than 50 $\mu$m. 
The invariant mass of these heavy groups should correspond to the $m_{h_2}$ and thus be larger than the bottom quark mass.
Thus, our next cut is the requirement to have at least one heavy group ($n_{\text{hG}}\geq1$) with an invariant mass above 6 GeV, which removes the background coming from hadronic decays of $B-$mesons. 

Then, we consider the inclusive invariant mass of all groups, $m_{SS}$.\footnote{At the parton level this quantity should equal the invariant mass of the two light scalars, hence the notation.} 
Na\"ively each heavy group should consist of two $b-$jets stemming from the decay of one $h_2$ and hence we expect the signal events to have two heavy groups with $m_{SS}$ peaking at the SM Higgs mass.
This dictates our last cut, which requires $m_{SS} \in [100,150]$ GeV.

\begin{table}[t]
	\small
\begin{center}
\begin{tabular}{c|c|c|c|c|c|c|c}
	cut & $n_J<5$ & $n_{\text{disp.} J}=0$ & $n_{\text{hG}}<1$ & $m_{SS}<100$ GeV & $m_{SS}>150$ GeV & $n_{\text{hG}}>2$ & $n_{\text{hG}}<2$   \\
\hline
	$B_1$ & 0.80 & 0.21 & 0.0012& $9.00$E-7 & $7.00$E-7  & $7.00$E-7 & $2.40$E-7  \\
	$B_2$ & 0.94 & 0.28 & 0.0020 & $3.20$E-6 & $2.58$E-6 & $2.58$E-6 & $4.60$E-7 \\
	$B_3$ & 0.98 & 0.33 & 0.0027 & $6.20$E-6 & $4.84$E-6 & $4.84$E-6 & $5.14$E-7 \\
	$B_4$ & 0.37 & 0.14 & 0.0013 & $4.48$E-6 & $3.70$E-6 & $3.70$E-6 & $3.00$E-7\\
	$B_5$ & 0.96 & 0.83 & 0.058  & $5.10$E-5 & $4.20$E-5 & $4.20$E-5 & $2.50$E-5\\
	$B_6$ & 0.99 & 0.85 & 0.062  & $1.16$E-4 & $9.10$E-5 & $9.10$E-5 & $4.60$E-5\\
	$B_7$ & 0.35 & 0.30 & 0.022  & $5.70$E-5  & $4.10$E-5 & $4.10$E-5 & $2.00$E-5\\
	$B_8$ & 1.00 & 0.96 & 0.17   & 0.0014 & 0.0010 & $9.90$E-4 & $7.50$E-4\\
	$B_9$ & 0.70 & 0.62 & 0.015  & $1.70$E-4 & $1.70$E-4 & $1.70$E-4 & 0.00\\
	$B_{10}$ & 0.87 & 0.76 & 0.02 & $3.20$E-4 & $2.40$E-4 & $2.40$E-4 & $1.00$E-5\\
	$B_{11}$ & 0.35 & 0.30 & 0.0085 & $1.50$E-4 & $1.30$E-4 & $1.30$E-4 & $2.00$E-5\\
	$B_{12}$ & 0.36 & 0.36 & 0.040  & 0.0011 & $7.90$E-4 & $7.90$E-4 & $1.00$E-4 \\
\hline\hline
	$10, 10^{-7}$ & 0.99  & 0.38  & 0.015 & $5$E-5 & $3$E-5 &  $3$E-5 & $1$E-5\\
	$10, 10^{-5}$ & 0.99  & 0.50  & 0.11  & 0.0062  & 0.0061 & 0.0061  & 0.0012  \\
	$10, 10^{-3}$ & 0.99  & 0.99  & 0.63  & 0.14    & 0.14   & 0.14    & 0.13    \\
	$10, 10^{-1}$ & 0.96  & 0.94  & 0.51  & 0.075   & 0.075  & 0.075   & 0.075   \\
	$10, 10^{1} $  & 0.61  & 0.15  & 0.018 & $8$E5 &  $8$E-5 &  $8$E-5 &  $8$E-5  \\
	$12, 10^{-7}$&  0.95 & 0.95 & 0.84 & 0.38 & 0.38 & 0.38  &  0.36\\
        $12, 10^{-5}$ & 0.95  & 0.95  & 0.86  & 0.41    & 0.41   & 0.41    & 0.39    \\
        $12, 10^{-3}$ & 0.95  & 0.95  & 0.87  & 0.45    & 0.45   & 0.45    & 0.44    \\
        $12, 10^{-1}$ & 0.86  & 0.85  & 0.69  & 0.27    & 0.27   & 0.27    & 0.27    \\
	$12, 10^{1} $  & 0.49  & 0.11  & 0.019 & $1.7$E-4 & $1.7$E-4 & $1.7$E-4 & $1.7$E-4 \\
	$30, 10^{-7}$ & 1.00  & 0.96  & 0.15  & $6.0$E-4 & $5.5$E-4  & $5.0$E-4 &  $4.1$E-4\\
        $30, 10^{-5}$ & 1.00  & 0.97  & 0.27  & 0.0044  & 0.0042 & 0.0041  & 0.0028  \\
        $30, 10^{-3}$ & 1.00  & 1.00 & 0.84  & 0.11    & 0.11   & 0.10    & 0.10    \\
        $30, 10^{-1}$ & 0.99  & 0.99  & 0.79  & 0.10    & 0.10   & 0.097   & 0.096   \\
	$30, 10^{1} $  & 0.54  & 0.17  & 0.048 & $2.6$E-4 & $2.4$E-4 & $2.3$E-4 & $2.3$E-4 \\
	$50, 10^{-7}$  & 1.00  & 0.97  & 0.17  & $2.5$E-4 & $2.2$E-4 & $1.7$E-4 & $1.4$E-4 \\
        $50, 10^{-5}$ & 1.00  & 0.98  & 0.19   & $6.4$E-4 & $5.8$E-4 & $5.0$E-4 & $3.5$E-4 \\
        $50, 10^{-3}$ & 1.00  & 1.00  & 0.88  & 0.061   & 0.060  & 0.049   & 0.045   \\ 
        $50, 10^{-1}$ & 1.00  & 1.00  & 0.87  & 0.046   & 0.046  & 0.036   & 0.034   \\
	$50, 10^{1} $  & 0.58  & 0.25  & 0.10  & $1.8$E-4 & $1.5$E-4 & $1.1$E-4 & $1.1$E-4  
\end{tabular}
	\caption{Cutflow efficiencies for the background processes ($\epsilon^{\text{cut}}_{B_i}$) and a selection of signal benchmark points ($\epsilon_S^{\text{cut}}$).
		 The upper set is for the 12 background processes and the lower set is for signal benchmark points with combinations of $m_{h_2}$ in GeV and $c\tau$ in meter.
		 The scientific notation is used for cutflow efficiencies smaller than 0.001.
		 }
		\label{tab:cutflow}
	\end{center}
\end{table}

In table~\ref{tab:cutflow}, we list the cutflow efficiency of signal events for a set of benchmark points for different values of $m_{h2}$ and $c\tau$, and of the background processes discussed in the last subsection. We find that for our cutflow only the processes $B_1 - B_7$ would yield background events at the LHeC, while the rest would not contribute. We also note that the signal events have larger efficiencies for lifetimes in the range of $10^{-3} - 10^{-1}$ m. This comes at no surprise, given the considered spatial dimensions of the detector, its spatial resolution, and the kinematics of the process.

The expected number of signal events $N_S$ at the LHeC is given by
\begin{eqnarray}
\label{eq:ns}
	N_S = N_{h_1} \cdot \text{Br}(h_1\rightarrow h_2 h_2)\cdot \big(\text{Br}(h_2 \rightarrow b \bar{b})\big)^2 \cdot \epsilon^{\text{pr-cut-XS}}   \cdot \epsilon^{\text{cut}}_{S}, \label{eqn:NS}
\end{eqnarray}
where $N_{h_1}=1.1\times 10^5$ is the total number of the SM Higgs bosons inclusively produced at the LHeC with a total integrated luminosity $\mathcal{L}_{\text{LHeC}}= 1$ ab$^{-1}$ \cite{Agostini:2020fmq}, $\epsilon^{\text{pr-cut-XS}}$ denotes the reduction of the signature production cross section from the generator-level cuts, and $\epsilon^{\text{cut}}_{S}$ comprises the detector acceptance and the final cutflow efficiency of the signal events.
Analogously, the total number of background events $N_B$ is simply
\begin{eqnarray}
	N_B = \sum_{i=1}^{12} \mathcal{L}_{\text{LHeC}}\cdot \sigma_{B_i}\cdot \epsilon^{\text{cut}}_{B_i}, \label{eqn:NB}
\end{eqnarray}
where $\sigma_{B_i}$ denotes the parton-level cross section of the $i-$th background processes, and $\epsilon^{\text{cut}}_{B_i}$ is the equivalent of  $\epsilon^{\text{cut}}_{S}$ for $B_i$.

As we described in the previous section, the first four background processes have very large cross sections, and due to computational limitations their associated statistical uncertainties are substantial. 
In order to make a more robust prediction of the expected number of $B_1 - B_4$ processes we perform a fit using the $m_{SS}$ sidebands.
Concretely, we select from each process the sample passing the $n_{\text{hG}}=2$ cut, and fit the $m_{SS}$ distribution outside the $[100,150]$ GeV Higgs window with an exponential function. 
We have validated this procedure by applying it to the other background processes $B_5 - B_{12}$, finding excellent agreement between the simulated and predicted numbers of events in the signal region. Combining this fitting procedure with the information given in table~\ref{tab:bgd} and table~\ref{tab:cutflow} we find the total number of expected background events at the LHeC to be $N_B = 195$.

Before closing this section, several comments are in order.
First, we have found that the efficiency of our cutflow, for a fixed $c \tau$, degrades with $m_{h_2}$. Indeed, for $m_{h_2} \gtrsim $ 30 GeV we have found a large fraction of events with $n_{\text{hG}}=1$ and $m_{SS} < 100$ GeV, while for $m_{h_2} \in [10-20]$ GeV most of the signal events pass the final cut. The reason behind this behavior is readily understood: a light $m_{h_2}$ would have more collimated $b-$quarks, and hence the decay products from this $h_2$ scalar are more likely to form a `heavy group' as defined above, and $m_{SS}$ is simply $m_{h_2 h_2}$. 
For larger values of $m_{h_2}$ each $b-$quark is more likely to form its own `heavy group', being efficiently removed by the 6 GeV invariant mass cut that aims at removing $B-$meson decays. 
Hence, it is likely that one or more $b-$quarks do not contribute to $m_{SS}$, which then peaks at lower values. 
As a matter of fact, the goal of our cut-based analysis was the light $h_2$ regime, where ATLAS and CMS fail to be competitive due to the large $H_T$ trigger used for displaced jet searches~\cite{Csaki:2015fba,Aad:2019xav,CMS:2020idp}.

Second, we would like to assess the impact of our strategy in the prompt regime of $h_2$  ($c\tau\lsim 10^{-6}$ m). In that case, the lifetime of $h_2$ is shorter than the natural lifetime of the $B-$mesons originating from the $b-$quark hadronization. Our search strategy targets then the $B-$decays, and hence the cutflow does not depend on the actual value of the $h_2$ lifetime.  Once in the prompt regime, lighter masses have larger efficiencies \footnote{An exception arises when $h_2$ is close to the kinematical threshold $\sim 10$ GeV.}, for reasons explained in the previous paragraph.
We note that a prompt search strategy for such light scalars decaying into a pair of $b-$quarks at the LHeC has been performed in ref.~\cite{liu:2016ahc}, obtaining an expected 6\% lower bound on the exotic Higgs branching fraction for an integrated luminosity of 1 ab$^{-1}$.

Third, we note that while the simple analysis depicted here can be improved, we have found no obvious additional handles to include. 
We have explicitly explored the use of the transverse missing energy, of including $n_{\text{hG}}=1$ or 3, and/or of further tightening the transverse displacement of the heavy groups. 
None of these options have led to a significant enhancement of the exclusion limits. 

Finally, we would like to comment on the choice of the $m_{SS}$  mass window, currently set between 100 and 150 GeV. 
We have explored the possibility to enlarge the mass window to include lower values, which enhances the number of signal events (particularly for $m_{h_2} \gsim 30$ GeV). Clearly, this also increases the background, and we found that the final exclusion limits do not change qualitatively when extending the mass window to $\sim 70$ GeV.

\section{Results}
\label{sec:resu}

In this section we present the sensitivity of our proposed search in a) the Higgs Portal model, introduced in section~\ref{sec:model} and b) a model-independent parametrization (using the  $c\tau$ -- Br$(h_1\rightarrow h_2 h_2)$ and $m_{h_2}$ -- Br$(h_1 \rightarrow h_2 h_2)$ planes). 
For the former we choose two independent values of the dark vacuum expectation value $x={10,100}$ GeV and display the results in the  $m_{h_2}$ -- $\sin^2 \alpha$ plane, as presented for instance in reference~\cite{Beacham:2019nyx}.

In the following we consider the previously derived number of background events $N_B = 195$ as a conservative estimate.
It is conceivable that our analysis strategy can be further refined in order to reduce the number of background events, for instance with an efficient generation of jet-merged samples, a jet-substructure analysis to discriminate the collimated $h_2 \to b \bar{b}$ decays present in the signal which are absent in the SM backgrounds, or by employing deep-learning algorithms.
Subsequently, we shall also consider the optimistic case of zero background, $N_B = 0$.

\subsection{Higgs Portal model results}
\begin{figure}[t]
\centering
\includegraphics[width=0.5\textwidth]{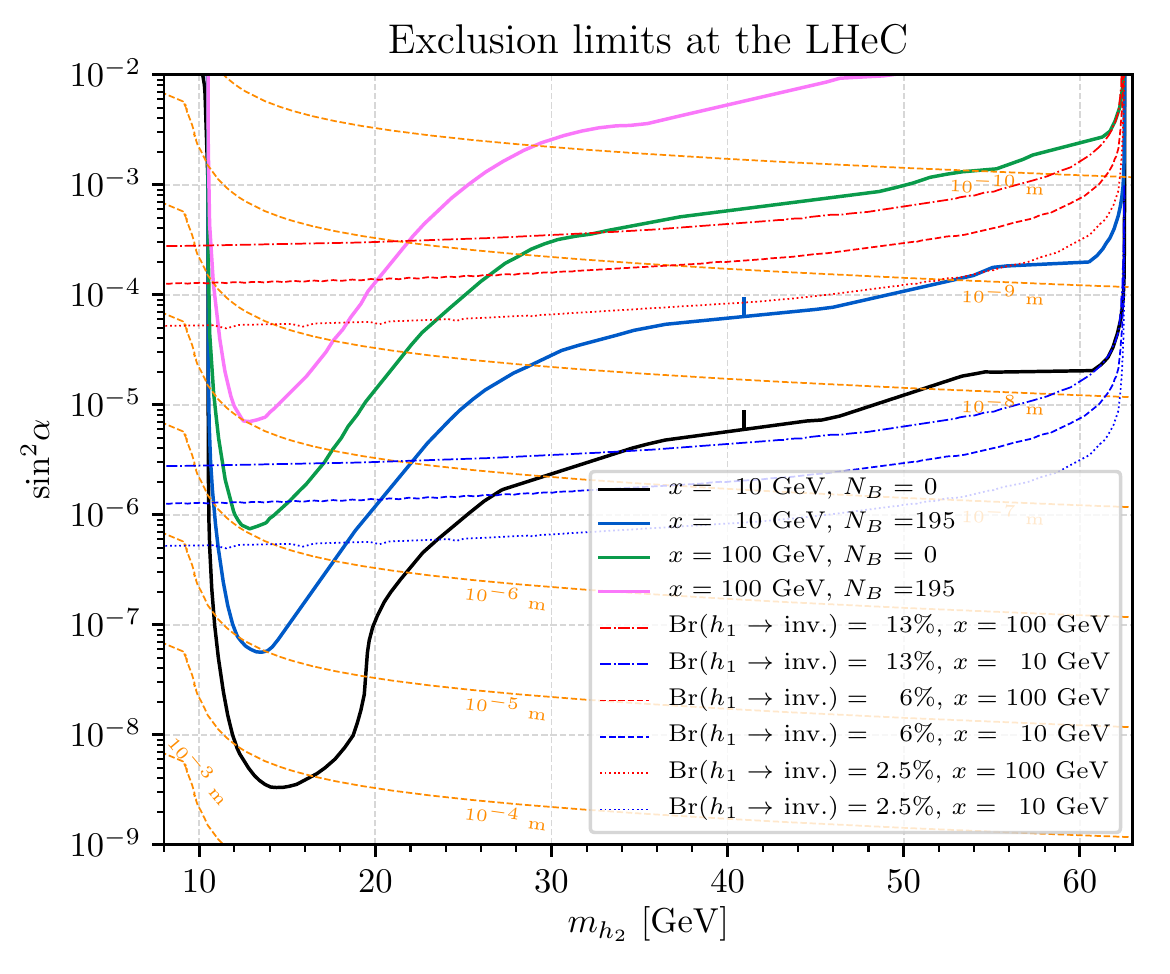}
	\caption{Model-dependent results shown in the plane $\sin^2{\alpha}$ vs. $m_{h_2}$.
	Two benchmark values of $x$ are selected, and the lifetime of $h_2$ is also plotted with orange dashed curves.
	 We superimpose the following limits on the Higgs invisible branching fraction: a) the present LHC $13\%$ ATLAS result~\cite{ATLAS:2020cjb} (dot-dashed); b) the expected outcome ($6 \%$) of a prompt LHeC search for these light scalars~\cite{liu:2016ahc} (dashed); c) the existing HL-LHC projection~\cite{Cepeda:2019klc} ($2.5 \%)$ (dotted). For these limits, blue (red) is for $x=10$ $(100)$ GeV.
 As discussed in the main text, direct current and future searches at the LHC for displaced jets~\cite{Aad:2019xav,CMS:2020idp} do not impact on the parameter space of this model.}
\label{fig:model-dependent}
\end{figure}
In this subsection, we present our results as a function of the parameters of the specific model introduced in section \ref{sec:model}, in particular we choose the mass-mixing squared plane.
The new vev, $x$, controls (for fixed scalar mixing and mass) the $h_1\to h_2 h_2$ branching ratio: as is shown in eq.~\eqref{eq:Higgs_branching}, the $h_1 \to h_2 h_2$ partial width scales as $x^{-2}$, and therefore a decrease in $x$ by a factor 10 corresponds to an enhancement in the partial width by a factor 100, while keeping other parameters fixed.\footnote{Since the partial width $\Gamma (h_1 \to h_2 h_2)$ is constrained to be much smaller than the SM Higgs width, $\Gamma (h_1 \to h_2 h_2) << \Gamma (h_1 \to {\rm SM}) \simeq 4.07$ MeV, the exotic partial width and the exotic branching fraction scale in the same way.}
In addition, $x$ also determines the mass scale of the dark sector. 
Hence, in order to avoid fine tuning in the scalar sector, we consider $x = {\cal O}(m_{h_2})$, and we fix the two representative values of $x = 10$ and 100 GeV for the Higgs portal model. 
Perturbative unitarity implies the constraint $m_{h_2} < \sqrt{16 \pi / 3}~x$ ~\cite{Robens:2015gla}, which correspondingly leads to an upper limit on $m_{h_2} < 41$ GeV for $x=10$ GeV, while for $x=100$ GeV this constraint does not affect the mass range considered in this work.\footnote{We thank Tania Robens for reminding us of this bound.}
In addition, we have explicitly checked that all quartic couplings remain perturbative, i.e: $ | \lambda_{1,2,3} | < 4 \pi$.

We show the resulting exclusion limits in the parameter plane $\sin^2{\alpha}$ vs. $m_{h_2}$ in figure \ref{fig:model-dependent}.
In this figure we use $N_B = 195$, as obtained from our MC study, and the optimistic assumption $N_B = 0$. 
For the case with $N_B=195$, we consider a significance level of $2 \sigma$ (or a 95 \% confidence level (C.L.)) with $N_S=2\,\sqrt{N_B}\sim 28$ where, given the magnitude of $N_B$, Gaussian statistics can be applied. 
As for the optimistic limits with $N_B=0$ we adopt Poisson statistics with a $95\%$ C.L. exclusion for $N_S=3$.
Moreover, we superimpose the current LHC \cite{ATLAS:2020cjb} and expected HL-LHC indirect limits \cite{Cepeda:2019klc}, together with the LHeC prompt reach \cite{liu:2016ahc}.
We also display isocontour lines for various values of $c\tau$ (in meters) using dashed orange lines.
We note that the LHC direct searches for displaced $b$-jets~\cite{Aad:2019xav,CMS:2020idp} do not constrain the here shown parameter space of the Higgs Portal.
These searches are sensitive to lifetimes $c \tau \gtrsim 1$ mm, which implies very small mixing angles: $\sin^2{\alpha} \lesssim 10^{-8}$ (cf.\ fig.~\ref{fig:model-dependent}). The corresponding exotic branching fraction is too small to yield an appreciable amount of signal events at the LHeC.\footnote{Taken as a benchmark point, for $m_{h_2}=10$ GeV, $c\tau=1$ mm, and $x=10$ GeV, Br$(h_1\to h_2h_2)\approx 10^{-4}$.} 
For smaller lifetimes the HL-LHC pays an exponential price to keep in signal events, which is not compensated by the linear scaling of the exotic branching fraction with $\sin^2{\alpha}$.

We find that for all four cases (two values of $x$ and two values of $N_B$) the displaced jet search proposed in this work may probe scalar mixing angles below the current and future LHC limits, for different ranges of $m_{h_2}$.
In particular, for the small value of the light scalar vacuum expectation value ($x=10$ GeV) and a background-free environment, $\sin^2{\alpha}$ can be excluded at below $10^{-8}$ for $m_{h_2}$ between 12 and 18 GeV.
In general, for $m_{h_2}>10$ GeV the present upper bounds from the invisible Higgs decay searches exclude $\sin^2 \alpha \gtrsim 10^{-4}$, while the here presented sensitivities can test $\sin^2 \alpha$ as small as $10^{-5} (10^{-7})$ for $x=100 (10)$ GeV, even exceeding the expected sensitivity of the indirect HL-LHC reach via the invisible Higgs branching ratio. 

\subsection{Model-independent results}
\begin{figure}[t]
\centering
\includegraphics[width=0.495\textwidth]{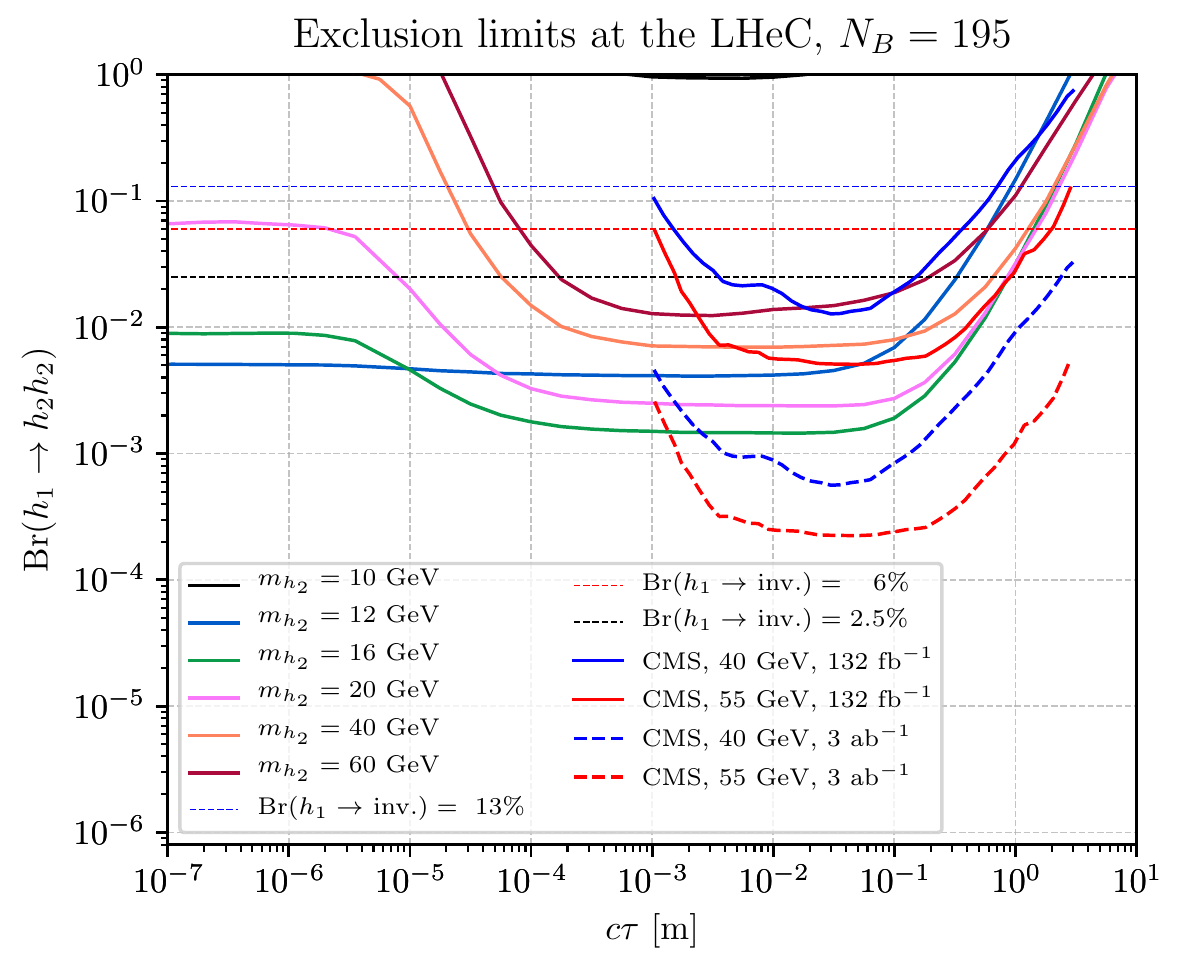}
\includegraphics[width=0.495\textwidth]{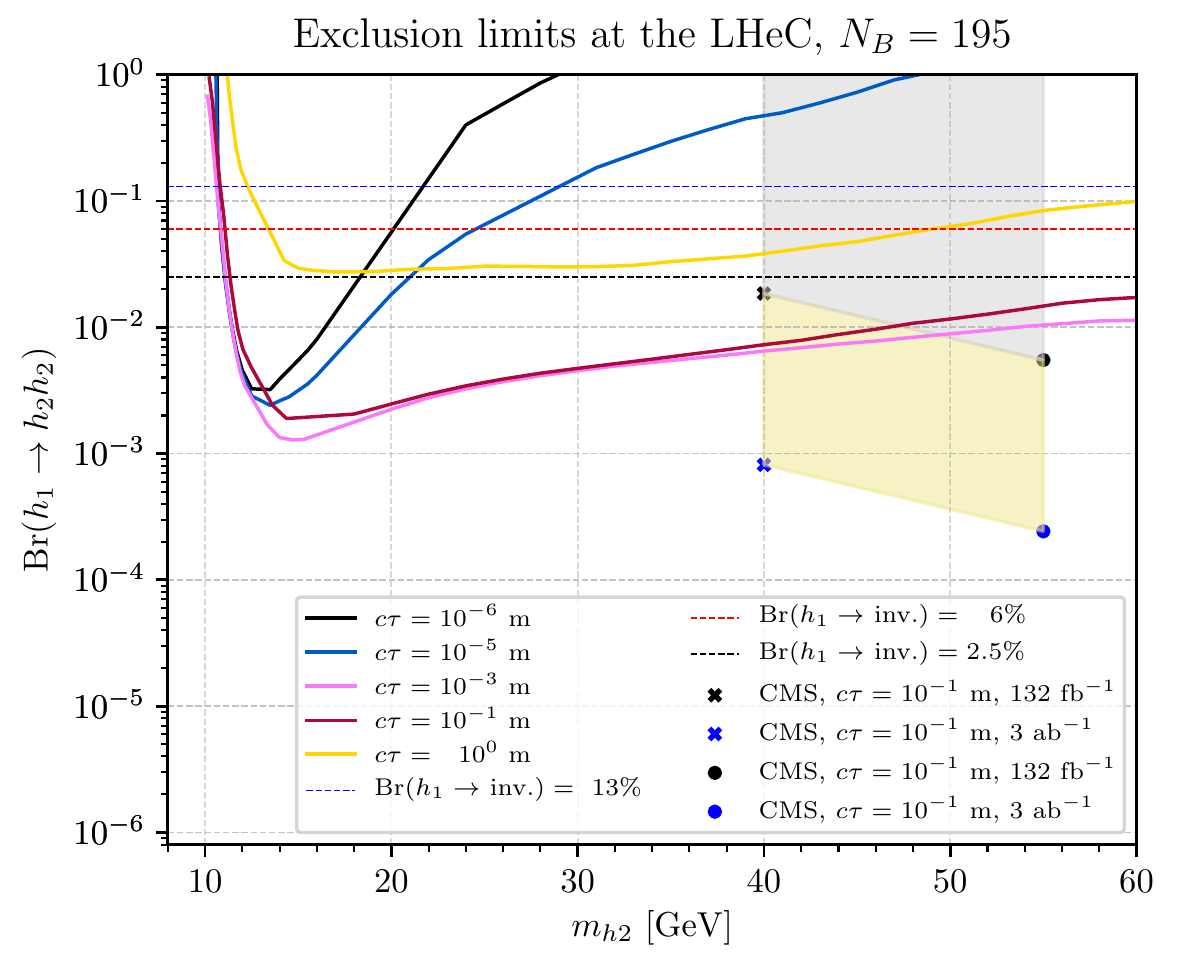}
	\caption{
	Sensitivity curves (95 \% C.L limits) in the $c\tau$ -- Br$(h_1\rightarrow h_2 h_2)$ (left) and $m_{h_2}$ -- Br$(h_1\rightarrow h_2 h_2)$ (right) planes. In the left (right) panel we display results for various values of  $m_{h_2}$ ($c\tau$). Here we have assumed SM-like branching fractions for $h_2 \to b \bar{b}$. 
	The present LHC, projected HL-LHC and future LHeC prompt search limits are also shown. See main text for details.
	We have assumed a total number of 195 background events at the LHeC with a total integrated luminosity of 1 ab$^{-1}$.
	In this figure we use the branching ratio of $h_2 \to b \bar{b}$ as shown in Appendix~\ref{app:hwidths}. 
	}
\label{fig:model-independent}
\end{figure}
In figure \ref{fig:model-independent}, we display our sensitivity estimations in a model-independent fashion in the Br$(h_1\rightarrow h_2 h_2)$ versus $c\tau$ and in the Br$(h_1\rightarrow h_2 h_2)$ versus $m_{h_2}$ planes. Here we have assumed a branching fraction of $h_2 \to b \bar{b}$ corresponding to a SM-like Higgs of the same mass, which approximately varies between 60-90\% in the considered mass range. For completeness, the current input values are presented in Appendix~\ref{app:hwidths}. 
Here we superimpose the bounds from the CMS search for displaced jets~\cite{CMS:2020idp} for the two benchmark masses chosen in the analysis: 40 and 55 GeV.

Furthermore, we na\"ively extrapolate the reach to the high-luminosity LHC by scaling the current limits with the total luminosity, assuming the searches remain background-free (currently there is $\sim 1$ expected background event in the signal region).%
\footnote{This extrapolation should be taken with a grain of salt: new strategies for analysis and trigger may improve the sensitivity beyond our na\"ive estimate, while the pile-up at the HL-LHC may add new backgrounds, and, as mentioned in footnote \ref{footnote3}, the $b\bar b$ final state always suffers from reconstruction efficiencies, in particular for tertiary vertices.}
The left panel of figure \ref{fig:model-independent} presents curves for various values of $m_{h_2}$, namely 10, 12, 16, 20, 40, and 60 GeV. 
As expected, due to the kinematic threshold effect the LHeC's sensitivity to $h_2$ with $m_{h_2} = 10$ GeV, denoted by the black line, is always weaker compared to the LHC's sensitivity via the invisible branching ratio.
The strongest reach in Br$(h_1\rightarrow h_2 h_2)$ is given by $m_{h_2}$ between 12 and 20 GeV, probing about $\mathcal{O}(10^{-3})$ on Br$(h_1\rightarrow h_2 h_2)$ for $c\tau$ between $10^{-4}$ and $10^{-1}$ m which is the most sensitive regime in accordance with the fiducial tracking volume.

For larger masses with a fixed $c \tau$, the sensitivity weakens due to the stronger $m_{SS}$ cut, as we anticipated in our discussion in section~\ref{sec:setup}. 
We note that in the large mass range the displaced jet searches from CMS~\cite{CMS:2020idp} are more sensitive than the LHeC estimates. Regrettably, CMS has only used benchmarks for the $h_2 \to b \bar{b}$ final state of 40 and 55 GeV, thus making it difficult to extrapolate their sensitivities for lower masses. Fortunately, they do present a 15 GeV benchmark for the case of $h_2$ decaying into light jets, and while they explicitly warn the reader, in Section 7.2, that the limits for such light masses in the $b \bar{b}$ final state are worse ``because the decays of b-hadrons can produce tertiary vertices, which can be missed by the secondary vertex reconstruction we deploy in this search". We will ignore this warning for the moment and conduct a back-of-the-envelope estimation of the sensitivity for 15 GeV. We start by observing that the ratio of the exclusion limits of $d \bar{d}$ over $b \bar{b}$ final states, for the two benchmark masses of 40 and 55 GeV, is relatively flat across their considered lifetime range, $c \tau \in [1 -10^{3}] $ mm, being about 0.1. The strongest limit for the 15 GeV benchmark of decays into light jets happens at $c \tau \sim 10$ mm, yielding a 8\% exclusion in the exotic branching fraction. Hence we can obtain the current (and HL-LHC) estimates for the $b \bar{b}$ final state, leading to an exclusion on the exotic branching fraction of $80\% (3.5\%)$ with current (HL-LHC) data via the direct search for light scalars decaying into $b \bar{b}$ pairs. 
With these assumptions for the LHC's sensitivity to a 15 GeV scalar we estimate that it requires a roughly 20 times improved sensitivity for the HL-LHC to be comparable to the LHeC's, while the latter (former) will remain more sensitive for masses below (above) about 30 GeV.

\begin{figure}[t]
\centering
\includegraphics[width=0.495\textwidth]{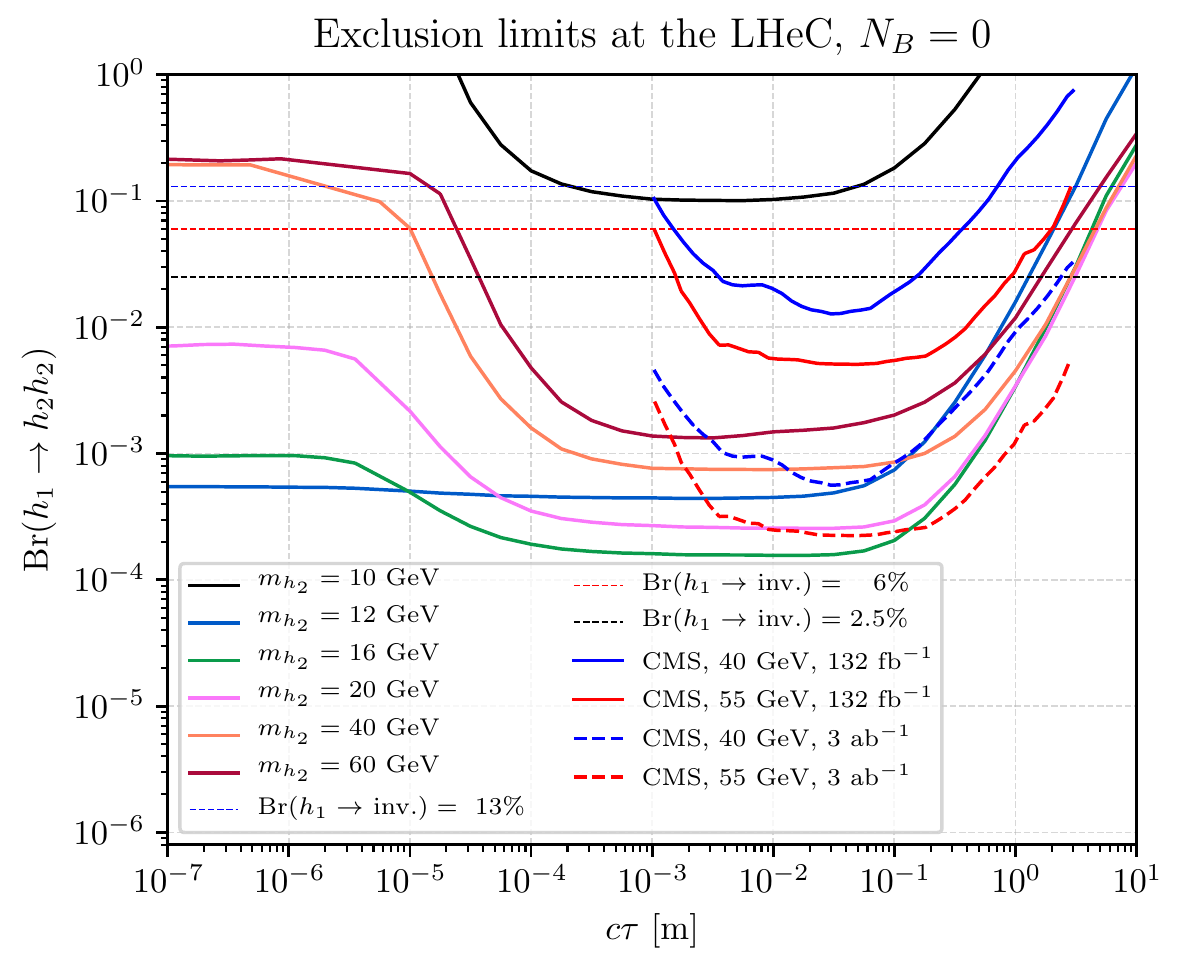}
\includegraphics[width=0.495\textwidth]{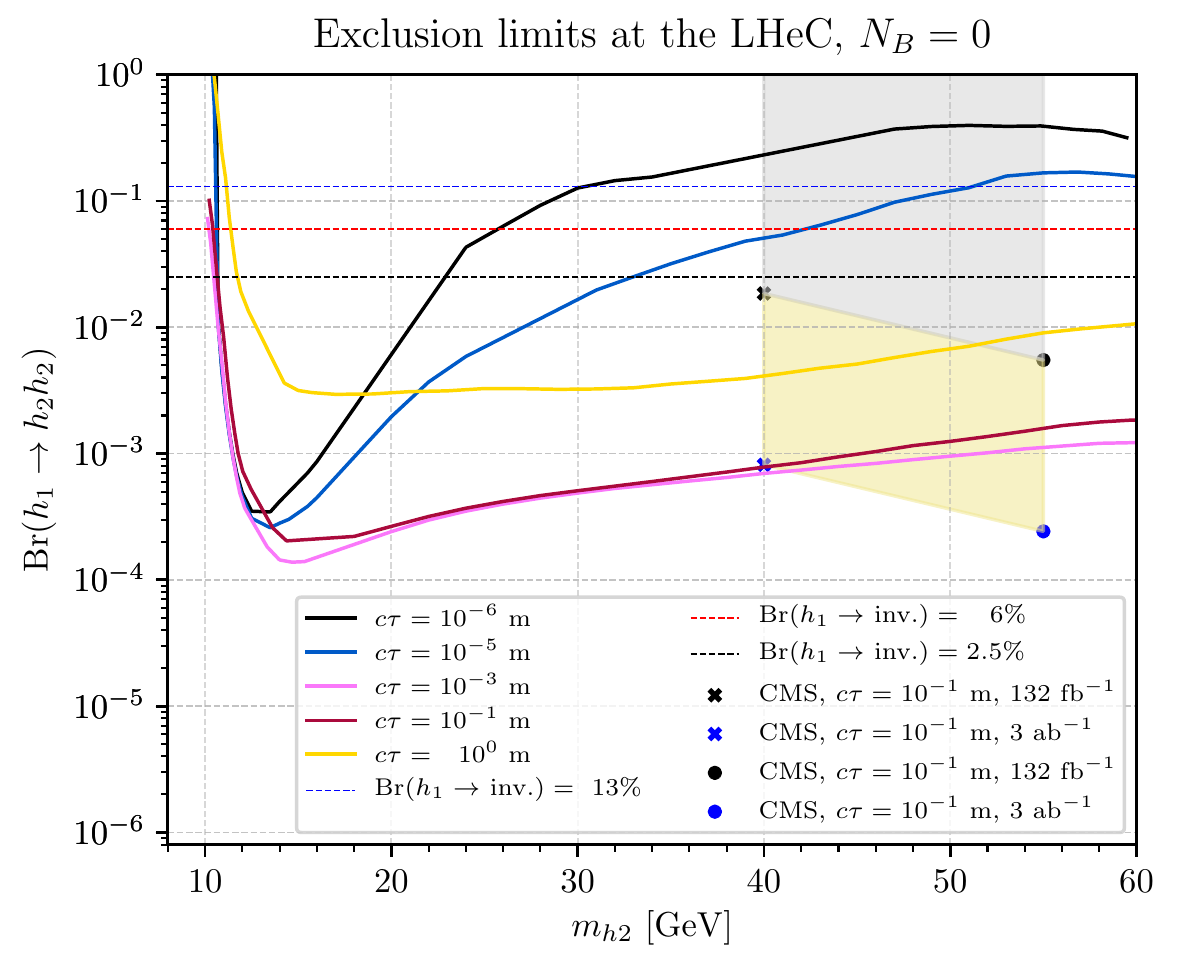}
	\caption{
	Same as figure \ref{fig:model-independent}, but for the ideal $N_B=0$ case.
	}
\label{fig:model-ind-no-bgd}
\end{figure}

For completeness we show the exclusion limits under the optimistic assumption of a background-free search in figure \ref{fig:model-ind-no-bgd}. 
In this case we adopt Poisson statistics with a 95\% C.L. exclusion for $N_S = 3$. 
This figure can be simply obtained by rescaling the $y-$axis of figure \ref{fig:model-independent} by a factor of $28 / 3 \sim 9.3$, since $N_S$ scales linearly with Br$(h_1\rightarrow h_2 h_2)$, as shown in eq.~\eqref{eq:ns}. Correspondingly, the maximum reach is increased by an order of magnitude, falling close to the $10^{-4}$ level.
Similarly, to obtain limits for other values of $N_S$ (e.g. for discovery prospects) one can simply rescale the $y-$axis of each plot.

We remark that for $c\tau \lesssim$ 1 $\mu$m the $h_2$ decays are practically prompt, proceeding essentially at the interaction point into $b-$jets.
In this regime the reconstructed displacement of the final state cannot be disentangled from the displaced decays of the $B-$mesons.
Therefore, the selection efficiencies are much smaller compared to the regime of long lifetimes and the corresponding sensitivity is finite (not vanishing) but weaker, as discussed above in section \ref{sec:setup}.
The resulting sensitivity of the LHeC to these exotic Higgs decay branching ratios in the prompt regime is 0.1 - 1\%, depending on the mass and background assumption, and independent of the $h_2$ lifetime.
This is in agreement with the analysis in ref.~\cite{liu:2016ahc}, where prompt $h_2$ decays were considered .

The right panel of figure \ref{fig:model-independent} exhibits the variation in the Br$(h_1\rightarrow h_2 h_2)$ reach as a function of the light scalar mass, with the lifetime fixed at different values: $10^{-6}$, $10^{-5}$, $10^{-3}$, $10^{-1}$, and $1$ m.
The plot shows that the best limit is reached for the curve with $c\tau=10^{-3}$ m, closely followed by the $c \tau=0.1$ m line. For larger $c \tau$ the light scalars decay mostly outside the inner tracker and are thus not available as tracks, resulting in a sensitivity that is similar to the HL-LHC projection, while for the prompt regime ($c \tau \lesssim 10^{-6} $ m) we observe that LHeC can place stronger bounds than the HL-LHC for light masses (11 GeV $\lesssim m_{h_2} \lesssim$ 15 GeV).

We remark that our analysis only makes use of tracks. 
We expect that including neutral LLPs decaying in other components of the detector should extend our sensitivity to $c\tau$ of up to a few meters, similar to what was done in the LHC analyses e.g.\ in refs.~\cite{Sirunyan:2017sbs,Aaboud:2019opc,Sirunyan:2019gut}. 
In particular, using calorimetric information could lead to very strong bounds for lifetimes above 100 meters, cf.\ ref.~\cite{Liu:2020vur}.

Summarizing, we find that searches for displaced jets at the LHeC can reach sensitivities to Br$(h_1\rightarrow h_2 h_2)$ as small as $\sim 10^{-3}$, which is about two orders of magnitude smaller than the  current LHC sensitivity on the Higgs invisible decay branching ratio ($13\%$)~\cite{ATLAS:2020cjb}, and also surpasses the expected HL-LHC performance of 2.5\% \cite{Cepeda:2019klc}. 
The direct searches for scalars decaying into a $b \bar{b}$ pairs conducted by ATLAS~\cite{Aad:2019xav} and CMS~\cite{CMS:2020idp} are more sensitive in the large mass regime and for $c \tau \gtrsim 1$ mm. For smaller masses and shorter lifetimes the LHC's sensitivity degrades rapidly with decreasing mass and lifetime. 
This mass and lifetime regime corresponds to the parameter region where th LHeC is most sensitive, thus complementing the reach of the HL-LHC searches.
We emphasize that, in contrast to the indirect search at the LHC, the LHeC search is in principle able to determine the masses and possibly also the lifetimes of the scalar LLPs.
%
\section{Conclusions}
\label{sec:conclu}

In this article we discussed the LHeC's sensitivity to the exotic Higgs branching ratio into scalar particles ($h_2$) via the search for displaced $b$-jets.
Existing constraints on the invisible Higgs branching fraction make it likely that the $h_2$ are long-lived, which renders the LHeC with its excellent spatial resolution and clean environment a natural setting for this study.

Our analysis includes the detailed simulation of background and signal events, accounting for parton shower and detector effects.
To deal with the spatial displacement from long-lived particles we employed the customized Delphes module from ref.~\cite{Nemevsek:2018bbt}. 
We emphasize that it would be highly desirable to incorporate this feature into existing fast detector simulation software.

We remark that many interesting scalar models featuring new phenomena exist which would non-trivially affect the sensitivity of the LHeC to $h_2$ decays. Examples are dark glueball or dark QCD models, where $h_2$ decays into semi-visible or invisible final states demand different search strategies. We leave this exciting venue for future work.
Here, we presented our results for a specific Higgs Portal model (SM + a complex singlet) and in a model-independent manner.
In the latter, the results were expressed in terms of the light scalar mass $m_{h_2}$, its lifetime $c \tau$, and the exotic branching fraction of $h_1$ into a pair of $h_2$ scalars, which allows one to readily generalize the results to other models.
Apart from presenting our results for the conservatively estimated number of background events ($N_B = 195$, obtained from our Monte Carlo study) we also considered, optimistically, the case of zero-background events.
For the concrete case of the Higgs portal, often used as a benchmark in the recent literature, the specific limit depends on the assumed value of the scalar vacuum expectation value (vev). 
We considered the new vev to be of the same order of magnitude as the new scalar mass: fixing two values 10 and 100 GeV in our analysis.
With this we find that the LHeC can test scalar mixings, $\sin^2 \alpha$, as small as $10^{-5} - 10^{-7}$ ( $10^{-6} - 10^{-8}$), for masses between 10 and 20 GeV under the conservative (optimistic) assumption on the number of background events, which is better than the limits of invisible Higgs decay search at the HL-LHC.

Regarding the model-independent limits, we found that the LHeC can cover parameter space currently untested by existing and future experiments, that is, scalar lifetimes between $10^{-4}$ and $10^{-3}$ m and branching fractions as small as $2 \times 10^{-3}$. Moreover, we expect that its sensitivity to longer lifetimes could be extended up to a few meters by including the calorimeters into the analysis, which is beyond the scope of the current work.
Searches for light scalars at the HL-LHC may be competitive with the 'indirect' search of an invisible Higgs branching fraction, which may test exotic branching fractions at 2.5\%.
HL-LHC searches for displaced $b-$jets may give even stronger limits for masses above $25-35$ GeV for $c \tau \in [10^{-3} - 1]$ m, but they may not be able to cover the range below about $25-35$ GeV, and/or for lifetimes below the millimeter.

It is important to realise that -- even in the case of a positive signal in the search for invisible Higgs decays -- the LHeC can be key in characterizing the signal by reconstructing the scalar masses and lifetimes.
All in all, we see that the LHeC will be able to reach a better sensitivity in both the low mass regime $[10-25]$ GeV and for lower $c \tau$ (probing with better accuracy lifetimes in the $10^{-5} - 10^{-3}$ m range, even under the assumption of a
substantial improvement of the sensitivity at the HL-LHC).
We thus conclude that it is worth pursuing a program of displaced-jet searches from the SM Higgs boson.

In summary, the sensitivity of the proposed LHeC search for light scalars coming from decays of the Higgs boson nicely complements those of dedicated experiments focusing on scalar masses below the $B-$meson mass, significantly extending the borders of the lifetime frontier into uncharted territory for masses above 10 GeV and lifetimes in the millimeter - meter range.

\begin{appendix}
 \section{Scalar widths and branching fractions}
 \label{app:hwidths}

\begin{figure}[t]
\centering
\includegraphics[width=0.495\textwidth]{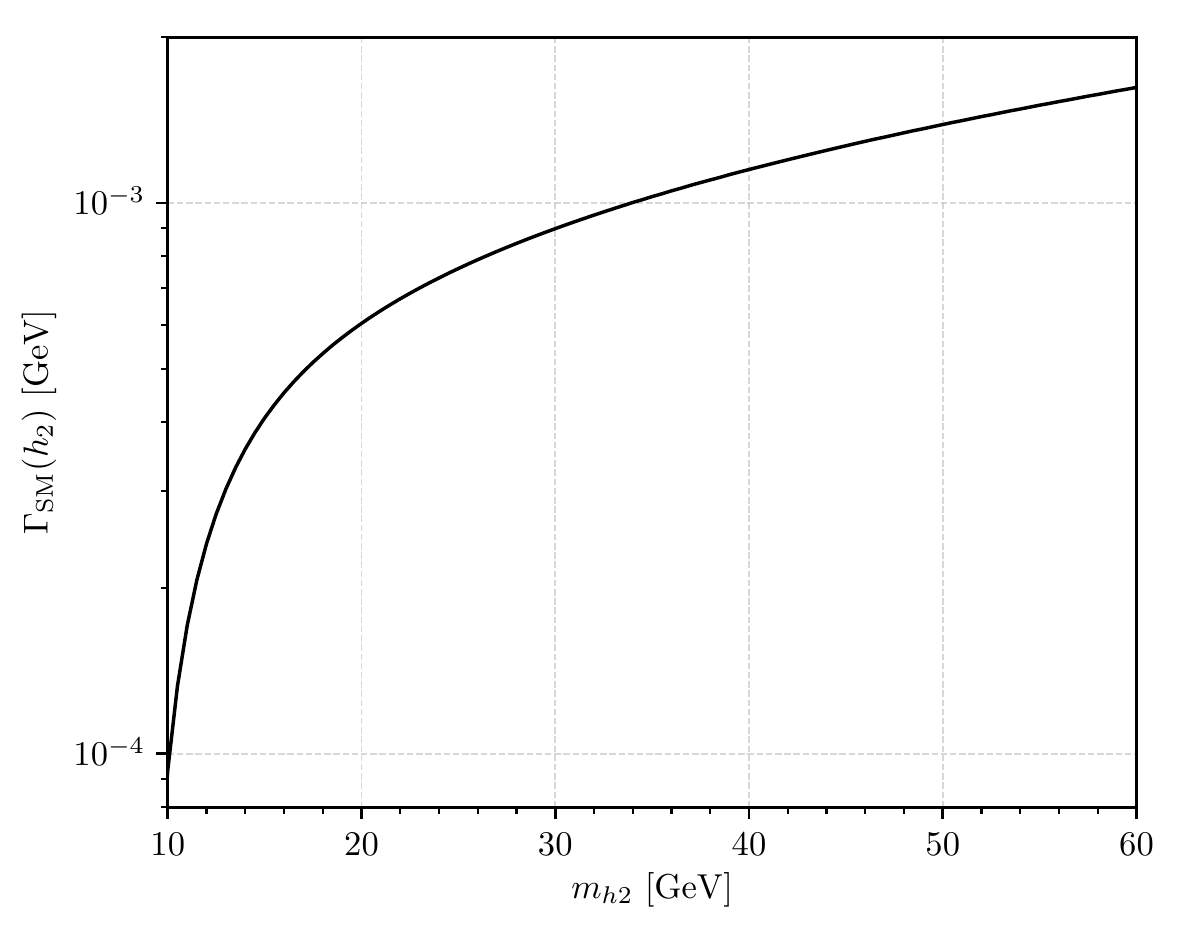}
\includegraphics[width=0.495\textwidth]{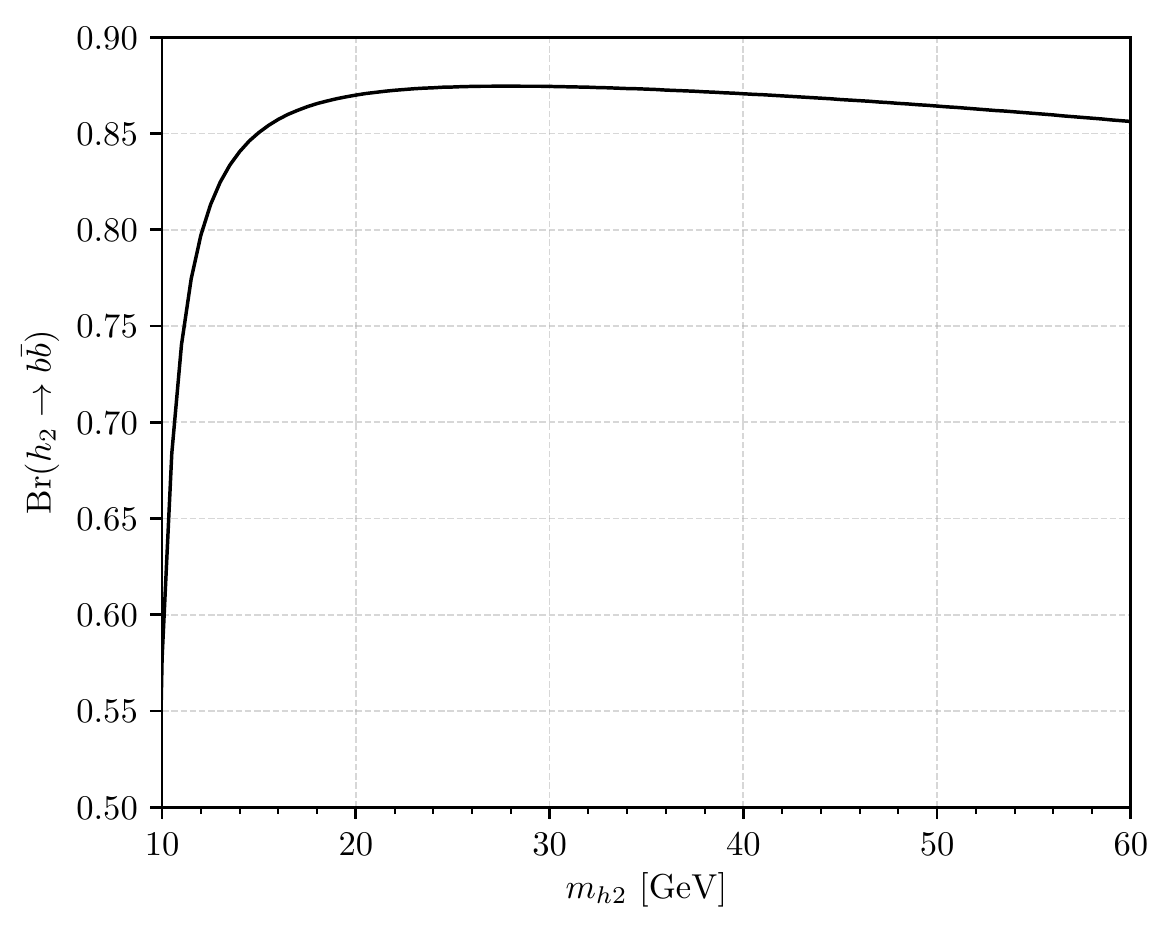}
	\caption{Input values for the total width (left) and the branching fraction into $b \bar{b}$ of the $h_2$, as a function of its mass. These values have been obtained by using HDECAY 3.4 \cite{Djouadi:1997yw,Djouadi:2018xqq}.
	}
\label{fig:app-hwidth}
\end{figure}

In this Appendix we present in figure~\ref{fig:app-hwidth}, for the sake of completeness and to facilitate the reinterpretation for arbitrary models, the input values assumed for the total width of the light scalar $h_2$ and for the branching fraction into $b \bar{b}$.
These were used to derive both, the model-dependent and the model-independent bounds presented in figures~\ref{fig:model-dependent},~\ref{fig:model-independent},~and~\ref{fig:model-ind-no-bgd}. 
These values allow the reinterpretation into other scenarios, where the $h_2$ widths (and branching fractions) can be different, for instance due to additional, exotic decays of $h_2$.

\end{appendix}

\section*{Acknowledgements:} 
We would like to thank Chen Zhang for collaboration in the early stages of this work. We thank Miha Nemev\v{s}ek for his detailed explanations on the use of the Delphes displaced module, and Xabier Cid Vidal, Nishita Desai, and Kechen Wang for useful discussions.
The authors acknowledge support from the LHeC Study Group.
Z.~S.~W. is supported partly by the Ministry of Science and Technology (MoST) of Taiwan with grant number MoST-109-2811-M-007-509, and partly by the Ministry of Science, ICT \& Future Planning of Korea, the Pohang City Government, and the Gyeongsangbuk-do Provincial Government through the Young Scientist Training Asia-Pacific Economic Cooperation program of the Asia Pacific Center for Theoretical Physics.
K.C. is partly supported by the MoST of Taiwan with grant no.
MoST-107-2112-M-007 -029 -MY3.

\bibliographystyle{JHEP}
\bibliography{EHD_ep,EHD_ep_v1}

\end{document}